\documentclass[twocolumn,floatfix,showpacs,prl,aps,tightenlines
]{revtex4}
\usepackage{graphicx}
\usepackage{psfrag}
\usepackage{dcolumn}
\usepackage{bm}

\newcommand{\bea}{\begin{eqnarray}}
\newcommand{\eea}{\end{eqnarray}}
\newcommand{\beq}{\begin{equation}}
\newcommand{\eeq}{\end{equation}}
\newcommand{\KMS}{\rm km\,s^{-1}}

\begin{document}

\title{Close encounters of three black holes}
\author{Manuela Campanelli}
\affiliation{Center for Computational Relativity and Gravitation,
School of Mathematical Sciences,
Rochester Institute of Technology, 78 Lomb Memorial Drive, Rochester,
 New York 14623}

\author{Carlos O. Lousto}
\affiliation{Center for Computational Relativity and Gravitation,
School of Mathematical Sciences,
Rochester Institute of Technology, 78 Lomb Memorial Drive, Rochester,
 New York 14623}

\author{Yosef Zlochower} 
\affiliation{Center for Computational Relativity and Gravitation,
School of Mathematical Sciences,
Rochester Institute of Technology, 78 Lomb Memorial Drive, Rochester,
 New York 14623}

\date{\today}

\begin{abstract} 

We present the first fully relativistic longterm numerical evolutions
of three equal-mass black holes in  a system consisting of a third
black hole in a close orbit about a black-hole binary.  We find that
these close-three-black-hole systems have very different merger
dynamics from black-hole binaries.  In particular, we see complex
trajectories, a redistribution of energy that can impart substantial
kicks to one of the holes, distinctive waveforms, and suppression of
the emitted gravitational radiation.  We evolve two such
configurations and find very different behaviors. In one configuration
the binary is quickly disrupted and the individual holes follow
complicated trajectories and merge with the third hole in rapid
succession, while in the other, the binary completes a half-orbit
before the initial merger of one of the members with the third black
hole, and the resulting two-black-hole system forms a highly
elliptical, well separated binary that shows no significant inspiral
for (at least) the first $t \sim 1000M$ of evolution.

 \end{abstract}

\pacs{04.25.Dm, 04.25.Nx, 04.30.Db, 04.70.Bw} \maketitle

{\it Introduction:}
The recent dramatic breakthroughs in the numerical techniques to evolve
black-hole-binary
spacetimes~\cite{Pretorius:2005gq,Campanelli:2005dd,Baker:2005vv} has
led to rapid advancements in our understanding of black-hole
physics. Notable among these advancements are developments in
mathematical relativity, including systems of PDEs and gauge
choices~\cite{Lindblom:2005qh,Gundlach:2006tw, vanMeter:2006vi},
the exploration of
the cosmic censorship~\cite{Campanelli:2006uy,Campanelli:2006fg,
Campanelli:2006fy,Rezzolla:2007rd,Sperhake:2007gu}, and the application of
isolated horizon
formulae~\cite{Ashtekar:2000hw,Dreyer:2002mx,
Schnetter:2006yt,Campanelli:2006fg,Campanelli:2006fy,Krishnan:2007pu}.
These breakthroughs have also influenced the development of data
analysis techniques
through the matching of post-Newtonian to fully-numerical
waveforms~\cite{Pan:2007nw,Boyle:2007ft,Hannam:2007ik}.
Similarly, the recent discovery of very large merger recoil kicks
\cite{Campanelli:2007ew,Gonzalez:2007hi,Campanelli:2007cga,
Herrmann:2007ac,Koppitz:2007ev,Herrmann:2006ks,Baker:2006vn,Gonzalez:2006md}
has had a great impact in
the astrophysical community, with several groups now seeking for
observational traces of such high speed holes as the byproduct of
galaxy collisions~\cite{Bonning:2007vt,HolleyBockelmann:2007eh}.
In this letter, we continue our quest to discover new astrophysical
consequences of black-hole interactions by simulating close
encounters of three black holes to see the different
behaviors introduced by the finite size of the holes, their nonlinear
interactions, and the radiation of
gravitational waves, as described by General Relativity.
We find that the three-body relativistic problem shows far richer
dynamics than the two-body problem, akin to the rich three-body
dynamics in Newtonian gravity, but with added complexity due to
mergers.

Three-body and four-body interactions are expected to be common in
globular clusters~\citep{Gultekin:2003xd,Miller:2002pg}, and in
galactic cores hosting supermassive black holes (when
stellar-mass-black-hole-binary systems interact with the Supermassive
black hole).
Hierarchical triplets of massive black holes might also be formed in
galactic nuclei undergoing sequential
mergers~\citep{Makino:1990zy,valtonen96}.  The gravitational
wave emission from such systems was recently estimated using post-Newtonian
techniques~\cite{Gultekin:2005fd}.

{\it Techniques:}
We evolve the three-black-hole
data-sets using the {\sc LazEv}~\cite{Zlochower:2005bj} implementation
of the `moving puncture approach'~\cite{Campanelli:2005dd,Baker:2005vv}.
We use the Carpet~\cite{Schnetter-etal-03b} driver to
provide a `moving boxes' style mesh refinement. In this approach
refined grids of fixed size are arranged about the coordinate centers
of each hole.
We use {\sc AHFinderDirect}~\cite{Thornburg2003:AH-finding} to locate
apparent horizons.
We extract the waveform on spheres centered about the origin 
and extrapolate the radiated energy/momentum to $r=\infty$
(the waveforms do no change qualitatively for
$r>50\pm20M$).

{\it Results:}
We chose one
configuration (3BH1) with purely {\em ad-hoc} momentum parameters, which
merged relatively quickly, to test the convergence and accuracy of our code. The
initial data parameters for these configurations are summarized in
Table~\ref{table:ID}. We evolved these configuration using 11 levels
of refinement and a finest resolution of $h=M/80$. The outer
boundaries were located at $640M$. In addition we evolved the 3BH1
configuration with grid-spacings rescaled by $5/6$ and $(5/6)^2$ to test
convergence. In the table, the horizon mass is the
Christodoulou mass, where ${m^H} = \sqrt{m_{\rm irr}^2 +
 S^2/(4 m_{\rm irr}^2)},$
$S$ is the magnitude of the spin of the hole, and $m_{\rm irr}$ is the
irreducible mass.
\begin{table}
\caption{Initial data parameters. $(x_i,y_i,0)$ and $(p^x_i, p^y_i,0)$ are
the initial position and momentum of the puncture $i$, $m^p_i$ is the puncture
mass parameter, and $m^H_i$ is the horizon mass.}
\begin{ruledtabular}
\begin{tabular}{lccc}
Config & 3BH1 &3BH101 & 3BH102 \\
\hline
$x_1/M$   & -2.40856   & -3.52462   & -3.52238   \\
$y_1/M$   &  2.23413   & 2.58509    & 2.58509    \\
$p^x_1/M$ & -0.0460284 & -0.0782693 & 0.0782693  \\
$p^y_1/M$ & -0.0126181 & -0.0400799 & -0.0433529 \\
$m^p_1/M$ & 0.315269   & 0.318143   & 0.317578   \\
$m^H_1/M$ & 0.335555   & 0.336201   & 0.335721   \\
$x_2/M$   & -2.40856   & -3.52462   & -3.52462   \\
$y_2/M$   & -2.10534   & -2.58509   & -2.58509   \\
$p^x_2/M$ & 0.130726   &  0.0782693 & -0.0782693  \\
$p^y_2/M$ & -0.0126181 & -0.0400799 & -0.0433529 \\
$m^p_2/M$ & 0.315269   & 0.318143   & 0.317578   \\
$m^H_2/M$ & 0.3405205  & 0.336241   & 0.335767   \\
$x_3/M$   & 4.8735     & 7.04923    & 7.04476    \\
$y_3/M$   & 0.0643941  & 0          & 0          \\
$p^x_3/M$ & -0.0846974 & 0          & 0          \\
$p^y_3/M$ & 0.0252361  & 0.0801597  & 0.0867057  \\
$m^p_3/M$ & 0.315269   & 0.320815   & 0.318585   \\
$m^H_3/M$ & 0.332198   & 0.333115   &0.331270    \\
\end{tabular} \label{table:ID} 
\end{ruledtabular} 
\end{table} 

It is interesting to note that the same techniques used for black-hole
binary evolutions work for configurations of three (and, according to 
a brief test by the authors, at least 22) black holes.
We tested the convergence of our algorithm with three
black holes by evolving configuration 3BH1 with three resolutions
($M/80, M/96, M/115.2$). We chose this configuration since it merges
relatively quickly, thus reducing the computational expense.
The resulting waveform converges to fourth-order. Note that
late-time accumulation of errors can have a significant effect on the
trajectories of 3-black-hole systems due to their inherent sensitivity to
changes in configuration. We have confirmed, through the use of a 
8th-order accurate code,
that the trajectories presented here show the correct qualitative behavior.

\begin{figure}
\begin{center}
\includegraphics[width=2.375in,height=1.9475in]{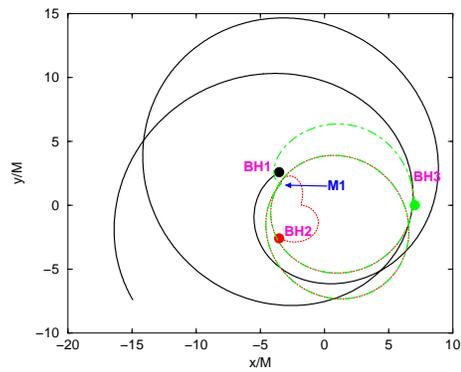}
\caption{The horizon trajectories for configuration 3BH101. 
The three black holes are initially located at the points labeled by BH1, BH2,
and BH3, respectively. BH1 and BH2 form a quasi-circular binary, which
is disrupted by BH3. BH2 and BH3 merge at point M1.
The BH1 and the BH2--BH3 merger remnant  continue to orbit
each other throughout the simulation.
}
\label{fig:cf101_track}
\end{center}
\end{figure}

\begin{figure}
\begin{center}
\includegraphics[width=3.0in,height=1.8in]{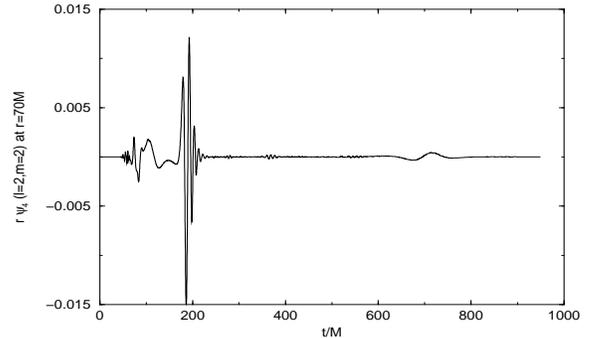}
\caption{The $(l=2,m=2)$ mode of $\psi_4$ for 3BH101. 
The BH2--BH3 merger waveform is centered at $t\sim185M$, the
small pulse at $t\sim700$ was produced by the close approach of
 the BH2--BH3 merger product to BH1.}
\label{fig:cf101_wave}
\end{center}
\end{figure}

We now show results for two similar initial configurations with qualitatively
different outcomes. To aid in the discussion, we will denote
the two holes in the binary with BH1 and BH2, and the third hole with
BH3, where BH1 is initially
located at $y>0$. 
We determined our initial data parameters by choosing a fiducial binary
configuration with orbital frequency $M_B\Omega_B = 0.04$ and angular
momentum $J_B/M_B^2 = 0.9104975$. We then treat the binary as a point
particle of mass $M_B = 2/3 M$ and spin $a_B/M_B = 0.9105$ along with
a non-spinning point particle of mass $M/3$, and choose position and
momenta parameters such that this two particle system is in a
quasi-circular orbit (up to 3 PN) at a separation equal to twice the binary's
separation. We set up the
systems so that the binary's orbital angular momentum is aligned with
the total orbital angular momentum (configuration 3BH101), and
anti-aligned (configuration 3BH102).
Configuration 3BH101 has BH2 and BH3 merging after
the binary completes nearly a half of an orbit. The result of this
interaction is to significantly push BH1 away from the merger remnant,
producing a new, highly elliptical, binary, with large orbital separation
oscillating in time from $10.4M$ to $23.5M$ (see
Fig.~\ref{fig:cf101_track}). 
The 3BH101 waveform (Fig.~\ref{fig:cf101_wave}) shows a burst of
radiation from the BH2--BH3 merger, as well as a small pulse at $t\sim
700M$ which corresponds to the point of closest approach of the
BH2--BH3 merger remnant with BH1. We stopped the evolution at $t\sim1000M$
due to computational expense and boundary contamination. 

\begin{figure}
\begin{center}
\includegraphics[width=2.375in,height=2.3275in]{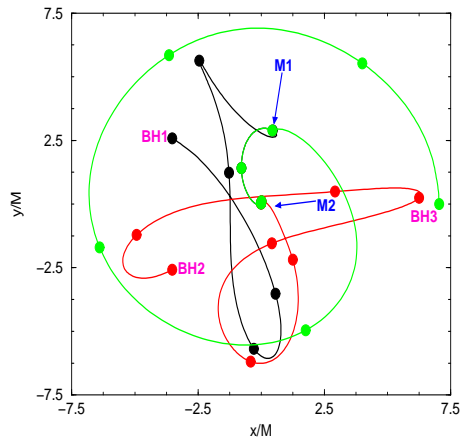}
\caption{The horizon trajectories for configuration 3BH102 with ticks every 
$45M$ of evolution.
The three black holes are initially located at the points labeled BH1, BH2,
and BH3, respectively. BH1 and BH2 form a quasi-circular binary, which
is almost immediately disrupted by BH3. BH1 and BH3 merge at point M1,
and then merge with BH2 at M2.
}
\label{fig:cf102_track}
\end{center}
\end{figure}

\begin{figure}
\begin{center}
\includegraphics[width=3.0in,height=1.55in]{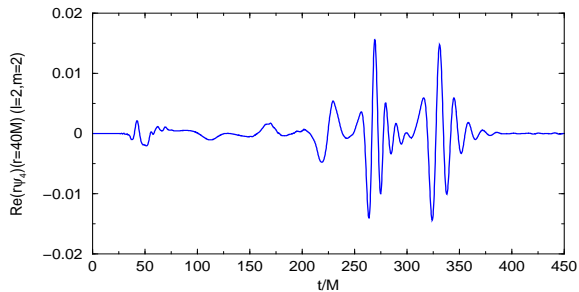}
\caption{The $(\ell=2,m=2)$ mode of $\psi_4$ for 3BH102 shows two
merger waveform signals arising from merger M1 and M2. }
\label{fig:cf102_wave}
\end{center}
\end{figure}

Configuration 3BH102 displays very different behavior, as seen in
Fig.~\ref{fig:cf102_track}. Here the binary is disrupted almost
immediately, and the individual holes follow complicated
trajectories (note that the trajectories are similar to the Greek
letters $\gamma$, $\tau$, and $\sigma$). BH3 and BH1 merge when BH3 almost
completes 1.25 orbits. The BH3--BH1 merger product then quickly merges
with BH2. The resulting waveform shows a double merger as seen in
Fig.~\ref{fig:cf102_wave}.
It is important to note that the dramatic difference in dynamics
between 3BH101 and 3BH102 is not strictly due to corotation versus
anti-corotation of the binary, but rather is a result the precise
configuration of the binary when the third black hole approaches.

\begin{table}
\caption{Total radiated energy, momentum, and angular momentum, as
well
as horizon mass spin and merger time, for 3BH102 and 3BH1.
($*$ 3BH101 did not merge in the time alloted and we only report
values for the first merger).}
\begin{ruledtabular}
\begin{tabular}{lccc}
Config & 3BH1 & 3BH101$^*$ & 3BH102 \\
\hline
$E_{\rm rad}/M (\times 1000)$    &$6.06\pm0.02$&$4.4\pm0.8$ &$6.1\pm0.7$\\
$J^z_{\rm rad}/M^2 (\times 100)$ &$2.92\pm0.01$&$3.6\pm0.5$&$4.4\pm0.2$\\
$P^x_{\rm rad}/M(\KMS)$          &$20.0\pm1.9$&$50\pm40$&$-1.3\pm14$\\
$P^y_{\rm rad}/M(\KMS)$          &$-22.9\pm2.4$ &$27\pm13$&$-15\pm13$\\
$M_{\rm H}/M$           &0.9835&***&0.9885\\
$S^z_{\rm H}/M^2$          &0.532&***& 0.465\\
$t_{\rm M1}/M$           &$\sim 27$&$\sim 115$&$\sim 218$\\
$t_{\rm M2}/M$                   &$\sim 40$&***&$\sim280$\\
\end{tabular} \label{table:Results} 
\end{ruledtabular} 
\end{table} 

For all three configurations the radiated energy and angular momenta
were a fraction of  that for a quasi-circular equal-mass
binary. This is due to the grazing type mergers, as also seen by the suppression
of the radiated angular momentum. We expect that the
full waveform for 3BH101 will show significantly more radiation as the
system eventually `circularizes' and merges.

We calculate the mass and spin of the first merger remnant and
the final remnant using the fitted exponential
decay rate and frequency of the quasi-normal behavior~\cite{Echeverria89}.
For 3BH102,
we fit the real and imaginary parts of the $(\ell=2,m=2)$ mode separately
and find $a_{H}/M_{H}=0.479, 0.378$ and $M_{H}=0.716, 0.678$ from fits
of the real and imaginary parts of $\psi_4$ for the first merger
product, and $a_{H}/M_{H}=0.478, 0.580$ and $M_{H}=0.994, 1.04$ from
fits
of the real and imaginary parts of $\psi_4$ for the  final merger remnant.
Note that the imaginary part of $\psi_4$ provides a better estimate
(i.e.\ closer to the expected $0.66$) mass for the first remnant, but the
real part provides a better estimate of both $a$ and $M_{H}$ for the
final remnant. We note the quasi-normal frequencies 
were $\omega = 0.64, 0.46$ for the first and final merger remnants respectively.
For 3BH101 we were only able to fit the real part of $\psi_4$ due
to significant contamination from other modes in the imaginary part.
We find a narrow region of width $10M$ where the waveform shows nearly
exponential decay. A fit to this region yields $\omega= 0.63$, with
a corresponding mass and spin of $M_{H} = 0.665$ and 
$a/M_{H} = 0.293$. The remnant masses, spins, and merger times
are given in Table~\ref{table:Results} for the 3BH1 and 3BH102
configurations.
The numbers quoted above should be taken only as indicative of the expected
values of these parameters. Further runs, at higher resolutions, will be
needed in order to establish the errors in these values.

{\it Discussion:}
 The relativistic  study of quasi-circular orbits of a binary in the
presence of a third comparable-mass hole, as an initial value problem,
was studied in~\cite{Campanelli:2005kr}.  Here, by studying their
dynamical evolution we find that the third
hole perturbs the system to the extent that no true binary orbit is
seen. We found that the close encounter of this third body can both
trigger a quick merger of the three-body system, as well as impart a
significant kick to one of the holes, producing a new long-lived,
highly-elliptical binary. The generic effect of the third black hole
is to reduce the gravitational radiation.  This happens for two
reasons. First, close-three-body interactions lead to grazing
collisions, which emit far less radiation than quasi-circular mergers.
Second, the resulting binary orbit will be elliptical, which is less
efficient at emitting gravitational radiation than circular orbits at
the final stages.  Note however, that although we report radiated
energies that are $1/{5^{\rm th}}$
that for a typical binary, here we scale the
energy by the total mass. If we scale the radiated energy with the
initial binary's mass, then the rescaled radiated energy would be 3/2
times larger.  The close-three-body systems also appear to be shorter
lived than typical binaries.

The three-black-hole waveforms (See Figs.~\ref{fig:cf101_wave}
and~\ref{fig:cf102_wave}) are distinct from the robust and simple form
of the binary-black-hole
waveform~\cite{Baker:2001nu,Baker:2002qf,Campanelli:2006gf,Baker:2006yw}.
In addition, there seems to be a large exchange of energy among the
components of the triple system, which occurs on a much shorter
timescale than the radiation.  It is important to note that these
three-black-hole interactions provide a mechanism for producing
highly-elliptical close-binaries, which would otherwise have
circularized (due to emission of gravitational radiation during the
inspiral).

Investigations of the Newtonian encounters of three bodies show that
such encounters generically lead to the breakup of the system into a
binary and the third body that escapes ~\cite{Monaghan76} in a
`water-shed effect'. The distribution of the eccentricity of the
remaining binary is bell shaped around $e=0.3$ for compact
systems~\cite{Duquennoy91,Kroupa95}.  Classical studies~\cite{Hills75}
show that the probability of exchange of the binary companion in a
triple system is surprisingly high for all comparable masses, reaching
near one for more massive $m_3$ \cite{SP:93}.  The motion of the
system can be chaotic, due to small denominators.  The finite size of
the black holes represents a natural regularization to the problem,
and the dissipative effects of the gravitational radiation can prevent
some configurations from becoming chaotic.

We have found that 3-black-hole systems exhibit complicated orbital
dynamics analogous to the rich 3-body Newtonian dynamics, but with the
added complexity introduced by mergers.  Further study of this
problem, including a comparison of Newtonian and relativistic
dynamics, will be reported in a forthcoming paper, where we also
examine configurations where the triple system is disrupted.

\acknowledgments
We thank Alessia Gualandris, David Merritt,
 and Hiroyki Nakano for valuable discussions.
We gratefully
acknowledge NSF for financial support from grant PHY-0722315,
PHY-0701566, PHY 0714388, and PHY 0722703; and NASA for financial
support from grant NASA 07-ATFP07-0158.
We also thank Hans-Peter Bischof for producing  3-D 
visualizations of the three-black-hole configurations introduced in this paper.
Computational resources were provided by the NewHorizons cluster at RIT and
the Lonestar cluster at TACC.

\bibliographystyle{apsrev}
\bibliography{../../../Lazarus/bibtex/references}

\end{document}